\documentclass[prb,twocolumn,superscriptaddress,floatfix,nopacs]{revtex4-1}
\usepackage{amsfonts,amsmath,epsf,dcolumn,natbib}
\usepackage{color}
\usepackage{graphicx}
\usepackage{multirow}
\usepackage{rotating}
\newcolumntype{.}{D{.}{.}{-1}}
\newcommand{\half}{\tfrac{1}{2}}

\newcommand{\be}{\begin{equation}}
\newcommand{\ee}{\end{equation}}
\newcommand{\bea}{\begin{eqnarray}}
\newcommand{\eea}{\end{eqnarray}}
\newcommand{\balg}{\begin{align}}
\newcommand{\ealg}{\end{align}}
\newcommand{\grad}{\ensuremath{\nabla}}
\newcommand{\ofr}{\ensuremath{(\mathbf{r})}}

\begin{document}

\title{Deorbitalization Strategies for meta-GGA Exchange-Correlation Functionals}
\author{Daniel Mejia-Rodriguez}
\email{dmejiarodriguez@ufl.edu}
\affiliation{Quantum Theory Project, Department of Physics, University of Florida, Gainesville, FL 32611}
\author{S.B.~Trickey}
\email{trickey@qtp.ufl.edu}
\affiliation{Quantum Theory Project, Departments of Physics and of Chemistry, University of Florida, Gainesville, FL 32611}

\date{11 October 2017}

\begin{abstract}
We explore the simplification of widely used meta-generalized-gradient
approximation (mGGA) exchange-correlation functionals to the Laplacian
level of refinement by use of approximate kinetic energy density
functionals (KEDFs). Such deorbitalization is motivated by the prospect of
reducing computational cost while recovering a strictly Kohn-Sham
local potential framework (rather than the usual generalized Kohn-Sham
treatment of mGGAs).  A KEDF that has been rather successful in
solid simulations proves to be inadequate for deorbitalization but we
produce other forms which, with parametrization to Kohn-Sham results
(not experimental data) on a small training set, yield rather good
results on standard molecular test sets when used to deorbitalize the
meta-GGA made very simple, TPSS, and SCAN functionals.  We also study
the difference between high-fidelity and best-performing
deorbitalizations and discuss possible implications for use in ab
initio molecular dynamics simulations of complicated condensed phase
systems.
\end{abstract}
                        
\maketitle

\section{\label{Intro}Introduction}

As is well known, the key ingredient for computational use of density functional
theory (DFT) in its Kohn-Sham (KS) form \cite{KS}  
is the exchange-correlation (XC) functional $E_{\mathrm {xc}}$.  It is 
the only term in the DFT variational functional
\begin{eqnarray}
E[n]&=&T_{\rm s}[n]+E_{\rm Ne}[n]+E_{\rm H}[n] \nonumber \\ [8pt]
&&+\,E_{\rm xc}[n] + E_{\rm NN},
\label{A2}
\end{eqnarray}
which requires approximation.  Here, as usual, for $N_e$ electrons 
in the field of fixed nuclei, 
the ground-state number density is $n({\mathbf r})$, 
the KS (non-interacting) kinetic energy functional is $T_{\rm s}[n]$,  
 the nuclear-electron interaction energy is $E_{\rm Ne}[n]$,  
 the Hartree energy (classical
electron-electron repulsion) is $E_{\rm H}[n]$, and the inter-nuclear 
repulsion energy is $E_{\rm NN}$.

Minimization in principle gives a single Euler equation.
Overwhelmingly predominant
practice however is to render the minimization in terms of the KS orbitals
via 
\be
n({\mathbf r}) = \sum_{i=1}^{N_e}\vert \varphi_i ({\mathbf r})\vert^2
\label{densdefn}
\ee
where for simplicity we have assumed unit occupation.  The 
non-interacting kinetic energy is 
\begin{eqnarray}
T_{\rm s}[\{\varphi_i\}_{i=1}^{N_e}]
&:=& \half \sum_{i=1}^{N_e}  \int\, d{\mathbf r} \, 
|\nabla \varphi_i({\mathbf r})|^2 \nonumber\\
&\equiv& \int d{\mathbf r} \, t^{\mathrm{orb}}_{\rm s}({\mathbf r}) 
\label{A1}
\end{eqnarray}
(the positive definite form in Hartree a.u.).  Minimization then gives 
the ubiquitous KS equation 
\be
\lbrace -\half \nabla^2 + v_{\mathrm{KS}}([n];{\mathbf r})\rbrace \varphi_i({\mathbf r}) = \epsilon_i \varphi_i ({\mathbf r}) \; .
\label{ordinaryKS}
\ee
Here $v_{\rm KS}=\delta (E_{\rm Ne} + E_{\rm H} + E_{\rm xc})/\delta n$ is the KS 
potential.  

For more than two decades, pursuit of accuracy combined with broad
applicability has led to increasingly complicated inclusion of
explicit orbital dependence in approximate XC functionals.  Such
inclusion occurs first on the meta-generalized-gradient-approximation
(mGGA)
\cite{PerdewEtAlMGGA99,ErnzerhofScuseria99,TPSSa,TPSSb,PRTCS07,TPRSCS07,ZhangVelaSalahub07,revTPSS,MVS2015,SCAN2015,TaoMo2016}
rung of the Perdew-Schmidt ladder of functional types
\cite{PerdewSchmidt01}.  mGGA functionals of this type use the KS
kinetic energy density $t^{\mathrm{orb}}_{\rm s}({\mathbf r})$ to
detect one- and two-electron regions. It is relevant to the present
work to note that another kind of mGGA depends on $\nabla^2 n$, not
$t^{\mathrm{orb}}_{\rm s}({\mathbf r})$.  For clarity, from here on we denote such
functionals as ``mGGA-L'' and reserve ``mGGA'' for those that use 
$t^{\mathrm{orb}}_{\rm s}({\mathbf r})$.

It has come to be accepted that the ratio \vspace{-4pt}
\be
\alpha[n]:= (t_{\mathrm s}^{\mathrm {orb}}[n]-t_W[n])/t_{TF}[n] := t_\theta/t_{TF}
\label{alphadefn}
\vspace{-2pt}
\ee
is the most useful form of KE-based region detector 
\cite{SunXiaoRuzsinszky12} in a constraint-based mGGA.   
Here the Thomas-Fermi \cite{Thomas,Fermi} and von Weizs\"acker \cite{Weizsacker}
KE densities are \vspace{-4pt} 
\bea
t_{TF} &=& c_{TF} n^{5/3}({\mathbf r}) \;\;\;, %
\;\;\;  c_{TF} :=  \frac{3}{10}(3\pi^2)^{2/3} \label{ttf} \\
t_W &=& \frac{5}{3} t_{TF} s^2 \label{tw}  \; ,
\eea
and \vspace{-4pt}
\be
s :=\frac{|\nabla n({\mathbf r})|}{2(3\pi^2)^{1/3}n^{4/3}({\mathbf r})} \; .
\label{sDefn}
\ee
Other dimensionless ratios that involve the kinetic energy density 
which appear in the mGGA context are
\be
z[n] := t_{W}[n]/t_s^{orb}[n]
\ee
and
\be
  w[n] := \frac{t_{TF}[n]/t_s^{orb}[n] - 1}{t_{TF}[n]/t_s^{orb}[n]+1} \; .
\ee
$w[n]$ has become very popular in the development of semi-empirical mGGAs \cite{Becke2000,MardirossianHeadGordon2017},
while $z[n]$ was used in such early non-empirical mGGAs as 
PKZB \cite{PerdewEtAlMGGA99} and TPSS \cite{TPSSa}.

The explicit orbital dependence of a mGGA introduced by $\alpha[n]$,
$z[n]$, and/or $w[n]$ makes the XC contribution to the KS potential,
$v_{\mathrm{xc}}=\delta E_{\mathrm{xc}}/\delta n$ accessible only via
the optimized effective potential procedure
\cite{StadeleMajewskiVoglGoerling1997,GraboKreibichGross1997,GraboKreibachKurthGross2000,HesselmannGoerling2008}.
On account of the computational complexity (hence cost) of the OEP
procedure that route rarely is taken in practice.  Instead,
calculations that use mGGAs typically also use the much less costly
generalized Kohn-Sham (gKS) procedure. It has a set of non-local
potentials $\delta E_{\mathrm{xc}}/\delta{\varphi_i}$ rather than the
local $v_{\mathrm{xc}}$.  The two schemes are not equivalent in
content \cite{YangPengSunPerdewGKSBandGap2016}.  Note also that 
gKS is not as efficient as an ordinary KS calculation with, for
example, a GGA functional.   In the context of DFT calculations to drive tens
of thousands of ab initio molecular dynamics
\cite{Barnett93,MarxHutter2000,Tse2002,MarxHutter2009} steps, the 
seemingly modest increment in computational cost might be quite significant.

The requirement that either OEP or gKS procedures be used 
arises from the explicit 
orbital dependence of the KS KE density in $\alpha[n]$, $z[n]$, or $w[n]$.
Thus it is at least
plausible that recent progress in so-called orbital-free DFT (OFDFT) might 
provide an alternative route for transforming
the mGGA problem into one with better computational efficiency and
with the interpretability of an ordinary KS equation.
The strategy is to evaluate $\alpha[n]$, for example, 
with an \emph{orbital-independent} approximation for $t_{\mathrm s}^{\mathrm {orb}}[n]$.  
So far as we know, that strategy has been studied only a few times previously.
First was with the LYP correlation functional \cite{LYP}.  
That had mixed success, partly because the Colle-Salvetti C functional 
\cite{Colle-Salvetti} that was deorbitalized has the peculiarity 
of being non-$N$-representable \cite{Morrison93} and partly because 
the TF \cite{Thomas,Fermi} kinetic energy density functional (KEDF)  
was used in the deorbitalizing role.  The other instance is Perdew and Constantin's  deorbitalization 
\cite{PerdewConstantin07} of the TPSS \cite{TPSSa} mGGA.  They used
a Laplacian-containing KEDF, showed that the result seemed to mimic 
the full TPSS mGGA, and gave a small sample of numerical results in 
support of that assertion. The two other deorbitalization 
examples of which we are aware \cite{JCP.142.154121,JCP.146.064105} 
addressed the non-additive part of subsystem DFT, hence lack direct
bearing on the original deorbitalization strategy. 

In the present work we take up 
various versions of the deorbitalization strategy.  We explore a 
wide variety of KEDFs, present an alternative KEDF parametrization  
scheme dependent on KS calculations only for a small set of atoms, and provide
thorough numerical validation 
in the context of standard molecular test sets.  In the next section we
outline the formulation and constraints, delineate the KEDFs 
we have considered, and summarize the code environment and techniques
implemented.  The third section gives parametrizations and comparative 
test results, followed by a brief concluding section.  

\section{\label{Form}Formulation, Constraints, and KEDFs}

The numerator $t_\theta$ in $\alpha$, Eq.\ (\ref{alphadefn}), is known in the orbital-free kinetic energy 
literature as
the ``Pauli term''.  It has two positivity properties 
\cite{PRB80,postNAMET} that are strict constraints on any $t_\theta$ 
approximation. They are 
\be
T_{\theta}[n] = \int d{\mathbf r} t_\theta[n({\mathbf r})] \ge 0
\label{Tthetapositive}
\ee
and
\begin{equation}
\frac{\delta T_{\theta}[n]}{\delta n({\bf r})} \ge 0 \; \forall {\textbf r} \; .
\label{vthetapositive}
\end{equation}
An important limitation is obvious.
Even if an approximate $t_\theta$ meets these 
constraints that does not ensure that it is a good local approximation,
That is, enforcement of (\ref{Tthetapositive}) and (\ref{vthetapositive})
does not guarantee that 
\be
t_\theta^{approx}[n] \approx t_s^{orb}[n] - t_W[n]
\label{tthetaconstraint}
\ee
for arbitrary proper $n$.  Nevertheless, progress on functionals 
$t_\theta^{approx}$ that do meet those constraints and do a reasonably
good job of reproducing conventional KS binding energy curves, 
equations of state, etc., \cite{JCAMD,PRB88,postNAMET} suggests that one should try those
functionals. 

To proceed it is useful to have definitions of other reduced
density derivatives (RDDs) in addition to the familiar reduced density
gradient $s$ defined at (\ref{sDefn}).
\begin{itemize}
	\item Reduced density Laplacian
	\begin{equation}
	q  := \frac{\nabla^2 n }{4(3\pi^2)^{2/3} n^{5/3} }
	\label{reduced-qdefn}
        \end{equation}
	\item Reduced density Hessian
	\begin{equation}
		p := \frac{\grad n \cdot \grad\grad n \cdot \grad n}{4(3\pi^2)^{2/3} \lvert \grad n \rvert^2 n^{5/3}}
	\label{reduced-pdefn}
        \end{equation}
	\item Reduced quadratic density Hessian
	\begin{equation}
		\tilde{p} :=  \frac{\grad n \cdot \grad\grad n \grad\grad n \cdot \grad n}{16(3\pi^2)^{4/3} \lvert \grad n \rvert^2 n^{10/3}}
        \label{tildep-defn}
	\end{equation}
\end{itemize}

From the great variety of available single-point KEDFs, initially  
we examined several to see if they reproduced $t_{\theta}[n]$ for densities 
from very accurate
Hartree-Fock (HF) orbitals for the first eighteen neutral atoms \cite{IJQC.71.491,Thakkar03}.
Those KEDFs tested include the second-order gradient expansion 
approximation (GEA2) 
\begin{equation}
t_{\textsc{gea2}}[n]  = t_{\textsc{tf}}[n]  + \tfrac{1}{9}t_{\textsc{w}}[n]  \;,
\label{GEA2}
\end{equation}
the Liu and Parr (LP) homogeneous functional expansion \cite{PRA.55.1792},
the APBEK functional \cite{PRL.106.186406},
the PBE2 and PBE4 functionals \cite{JCAMD}, VT84F \cite{PRB88},
Perdew and Constantin (PC) mGGA \cite{PerdewConstantin07}, 
Cancio and Redd (CR)
mGGA \cite {JCP.144.084107,CancioRedd2017}, the regularized -- by enforcing the von Weizs\"acker kinetic energy density lower-bound -- version of the
Thomas-Fermi plus Laplacian (TFLreg) mGGA \cite{JCP.146.064105}, 
and a modified version of VT84F 
\begin{equation}
t_{\textsc{mvt}}[n]  = \theta_{\textsc{mvt}}[n]  t_{\textsc{vt}}[n]  + \left( 1-\theta_{\textsc{mvt}}[n]  \right)t_{\textsc{w}}[n]  \; . 
\label{mvtdefn}
\end{equation} 
In it, $t_{\textsc{vt}}[n]$ is the original VT84F KEDF and $\theta_{\textsc{mvt}}[n]$ is
an interpolation function
\begin{equation}
	\theta_{\textsc{mvt}}  = \mathrm{Erf}\left[\sqrt{\Theta }\right]
\label{fbdefn}
\end{equation}
based on the Density Overlap Regions Indicator (DORI) \cite{JCTC.10.3745}
\begin{equation}
	\Theta := 4 \left( 1 + \frac{\tilde{p}}{s^4} - 2\frac{p}{s^2} \right)
\label{DORIdefn}
\end{equation}
Because $\Theta\ofr  = 0$ for the hydrogen atom and $\rightarrow \infty$
for the homogeneous electron gas (HEG) \cite{Chem.201603914}, use of 
$\theta_\textsc{mvt}$ with VT84F removes spurious isolated H contributions.  

The quality measure \cite{JCP.127.144109} we 
used to determine which functional might yield reasonable
approximations to $\alpha[n]$ was 
\begin{equation}
\sigma = \frac{\int d\mathbf{r} \; \lvert  t_{\theta}^{orb} - t_{\theta}^{approx}  \rvert }{T_s} = \frac{\int d\mathbf{r} \; \lvert  t_{s}^{orb} - t_{s}^{approx}  \rvert }{T_s} \;.
\label{quality1}
\end{equation}
\begin{table}
	\caption{Average $\sigma$ values for the first eighteen neutral atoms computed with several kinetic energy density functionals. ``Regularized'' denotes
conformance with the von Weizs\"acker lower bound. \label{sigma}}
	\begin{tabular}{l c c}
		Functional & Regularized? & $\sigma$ \\\toprule
		PBE2       & no           &  1.576        \\		
		VT84F      & no           &  1.405        \\
		PBE4       & no           &  1.272        \\
		LP         & no           &  1.112        \\
		APBEk      & no           &  1.028        \\
		TW02       & no           &  1.027        \\
		LC94       & no           &  1.027        \\
		OL2        & no           &  1.017        \\
		OL1        & no           &  1.016        \\
		GEA2       & no           &  1.013        \\
		E00        & no           &  0.996        \\
		LP+L       & yes          &  0.827        \\
		W          & yes          &  0.473        \\				
		RDA        & yes          &  0.382        \\
		CR         & yes          &  0.271        \\
		MVT84F     & yes          &  0.243        \\
		TW02+L     & yes          &  0.239        \\
		GEA2+L     & yes          &  0.237        \\
		MVT84F+L   & yes          & \textbf{0.164}  \\
		TFLreg     & yes          & \textbf{0.147}  \\
		PC         & yes          & \textbf{0.117}  \\
		CRloc      & yes          & \textbf{0.103}  \\\toprule
	\end{tabular}
\end{table}
 Numerical integrals were performed by 
double-exponential radial quadratures \cite{TCA.130.645} with
200 points. This numerical scheme matched, to machine-precision, the
integrated KE obtained by analytical integration for all eighteen atoms.

Table \ref{sigma} lists average $\sigma$ values for the
first eighteen neutral atoms. Four approximate KEDFs 
stand out: Perdew and Constantin (PC) mGGA \cite{PerdewConstantin07}, 
Cancio and Redd (CRloc)
mGGA \cite {CancioRedd2017}, the regularized version of the
Thomas-Fermi plus Laplacian (TFLreg) mGGA \cite{JCP.146.064105}, and
the modified VT84F plus Laplacian (MVT84F+L).
The PC mGGA uses a modified fourth-order gradient expansion (MGE4) with several appealing
features. Written in an enhancement function form,
\be
T_s[n] = \int\,d{\mathbf r}\, t_{TF}({\mathbf r})F_t(s,q,p) \; ,
\label{KEenhancement}
\ee
the MGE4 functional is 
\begin{equation}
F_t^{MGE4} = \frac{F_t^{(0)}+F_t^{(2)}+F_t^{(4)}}{\sqrt{1 + [F_t^{(4)}/(1 + F_t^{W})]^2}}
\end{equation}
with
\bea
F_t^{(0)}& =&  1 \\
F_t^{(2)} &=& \tfrac{5}{27} s^2 + \tfrac{20}{9} q \\
F_t^{(4)}& = &\tfrac{8}{81} q^2 - \tfrac{1}{9} s^2 q + \tfrac{8}{243} s^4  \;. 
\eea
Perdew and Constantin assumed that $F_t^{MGE4} < F_t^{W}$ indicates the need
for $F_t^{PC} = F_t^{W}$. Thus, their functional interpolates between
$F_t^{MGE4}$ and $F_t^{W}$ with 
\begin{equation}
\theta_{PC}(z) = \begin{cases}
0,                                                               & z \leq 0 \\
\left[ \frac{1 + e^{a/(a-z)}}{e^{a/z} + e^{a/(a-z)}} \right]^b,  & 0 < z < a\\
1,                                                               & z \geq a
\end{cases}
\end{equation} 
They optimized the parameters $a$ and $b$ by minimizing the mean absolute
relative error (MARE) of the integrated kinetic energy of 50 atoms and ions,
nine spherical jellium clusters (with bulk parameter corresponding to Na), and
three systems composed of eight jellium spheres calculated in the liquid drop
model.  The result was $a = 0.5389$ and $b = 3$. The final form of the PC mGGA
enhancement factor is
\begin{equation}
F_t^{PC} = F_t^{W} + z^{PC} \, \theta_{PC} \left(z^{PC}\right)
\end{equation}
with
\be
z^{PC} = F_t^{MGE4} - F_t^{W}
\ee

Cancio and Redd \cite{CancioRedd2017} noticed 
some odd behavior of the PC mGGA for regions of small $p$ and negative $q$
which led them to suggest their CR mGGA 
\cite{CancioRedd2017,JCP.144.084107}. (Remark: Cancio and Redd's $p$ variable
is {\it not} the same as the one defined at Eq.\ (\ref{reduced-pdefn}) above.
Further their definition has an obvious typographical error; it should have
$n^2$ in its denominator, not $n$.)  
The original CR mGGA is based on the second-order gradient expansion including
the Laplacian term
\begin{equation}
F_t^{GEA2+L} = 1 + \tfrac{5}{27} s^2 + \tfrac{20}{9} q 
\end{equation}
and imposition of the von Weizs{\"a}cker lower bound via the 
interpolation function
\begin{equation}
\theta_{CR}(z) =  \left\lbrace 1 -\exp\left[ -1/\lvert z \rvert^a \right] \left[ 1 - H(z)\right]\right\rbrace^{1/a} 
\end{equation}
Here $H(z)$ is the Heaviside unit step function.
The resulting CR mGGA enhancement function is
\begin{equation}
F_t^{CR} = 1 + F_t^{W} + z^{CR} \theta_{CR}\left( z^{CR} \right)
\label{CRmGGAenhancement}
\end{equation}
with
\be
z^{CR} = F_t^{GEA2+L} - F_t^{W} - 1
\ee
and $a=4$. They also gave an alternative formulation of their
functional (CRloc) in which the local gradient expansion
\be
F_t^{GEAloc} = 1 - 0.275 s^2 + 2.895 q
\ee
is used in place of $F_t^{GEA2+L}$. This local expansion is only valid for
the nuclear region and is not expected, because of the sign of the $s^2$ term, 
to yield accurate integrated KEs by itself. Cancio and Redd discuss this
in detail but it is not an issue for our purposes. 

The TFLreg mGGA is based on the TFL enhancement function
\begin{equation}
F_t^{TFL} = 1 + \tfrac{20}{9}q
\label{basicTFL}
\end{equation}
augmented by imposition of the von Weizs{\"a}cker lower bound
\begin{equation}
F_t^{TFLreg} = \mathrm{max}\left( F_t^{TFL}, F_t^{vW} \right) \; .
\label{TFLreg}
\end{equation}

The MVT84F plus Laplacian functional is an extension of 
Equation (\ref{mvtdefn}) 
\be
t_{\textsc{mvt+l}}  = \theta_{\textsc{mvt}} \left( t_{\textsc{vt}} + \tfrac{1}{6} \nabla^2 n \right)  + \left( 1-\theta_{\textsc{mvt}}  \right)t_{\textsc{w}}  \; . 
\ee
Despite its good performance, we did not pursue it because of
difficulties in integrating the DORI
function with standard numerical techniques used
in many DFT codes. Those difficulties arise from the high-order 
spatial derivatives introduced by DORI and exacerbated by the use of
Gaussian basis sets.

That leaves three candidate KEDFs.
Additional numerical estimates of the deorbitalization performance
of those candidates were obtained from several additional error
indicators.
A global error indicator useful for all orbital-dependent mGGAs, 
\bea
\Delta_{\alpha} &=& \sum\limits_i^M \frac{1}{N_i} \int d\mathbf{r} n_i | \alpha_i^{orb} - \alpha_i^{approx} | \\
                &=& \sum\limits_i^M \frac{1}{N_i} \int d\mathbf{r} n_i \frac{| t_s^{orb}[n_i] - t_s^{approx}[n_i] |}{t_{TF} [n_i]} \; ,
\eea
was suggested in Ref.\ \onlinecite{JCP.146.064105}. 
Here, $M$ is the number of systems tested, 
$N_i$ is the number of electrons in the $i$th system and
$\alpha$ is as in Eq.\ (\ref{alphadefn}). Unfortunately, both 
$\sigma$ and $\Delta_{\alpha}$ tend to emphasize errors
in a particular region and not over the whole space.
$\sigma$ has a strong bias to the core region where
$t_s$ is at its maximum, whereas $\Delta_\alpha$ tends to
favor the tail region where $\alpha \rightarrow \infty$.
A more balanced indicator can be obtained by restricting the radial integration
in  Eq.\ (\ref{alphadefn})
to a small sphere around the nuclei, yielding
\begin{equation}
\Delta_\alpha^{near} = \sum\limits_i^M \frac{4\pi}{N_i^{near}} \int\limits_0^4 dr \; r^2 n_i | \alpha_i^{orb} - \alpha_i^{approx} | 
\end{equation}
where $N_i^{near}$ denotes the fraction of electrons inside a sphere of
radius 4 Bohr. The sphere radius was chosen to be sufficiently large to
enclose all standard C-, N- and O- single bonds \cite{CRCHandbook}, and at the same time be 
sufficiently small to avoid divergence of $\alpha$.

Post-scf $\Delta_{\alpha}$ and $\Delta_{\alpha}^{near}$  values were
obtained for the eighteen neutral atom test set. 
\begin{table}
	\caption{Error indicator $\Delta_{\alpha}$ values for various mGGA kinetic energy density functional approximations. \label{t:Dalpha}}
	\begin{tabular}{l l l l }
		       & \phantom{a} & \multicolumn{1}{c}{$\Delta_{\alpha}$} & \multicolumn{1}{c}{$\Delta_{\alpha}^{near}$} \\\toprule
		       
		PC     &             &     0.470565                          &  0.232595  \\
		CRloc  &             &     0.437642                          &  0.185015  \\
		TFLreg &             &     0.243281                          &  0.152519  \\\toprule
	\end{tabular}
\end{table}
Table \ref{t:Dalpha} shows $\Delta_{\alpha}$ and $\Delta_{\alpha}^{near}$ values 
for the PC, CR, and TFLreg mGGA functionals. On the assumption that
those two error estimates are good indicators, those results point
to the
TFLreg functional as the potentially best performer in deorbitalizing a
mGGA exchange-correlation functional, at least for those
mGGAs in which orbital-dependence arises solely from $\alpha$.

\section{Reparametrization of KEDF} 

Unfortunately, some functionals with low $\Delta_{\alpha}$ values may 
yield poor thermochemistry outcomes. (We discuss this below in the
context of Tables \ref{t:MVSdeorb}-\ref{t:SCANdeorb}). 
To compensate for possible mis-assessment by $\Delta_{\alpha}$ and/or
$\Delta_{\alpha}^{near}$, we reoptimized the parameters in all three
functionals so as to minimize
$\Delta_{\alpha}+\Delta_{\alpha}^{near}$.  Re-optimization was
done post-scf, again for the first 18 neutral atoms.  The reoptimized
functionals are named PCopt, CRopt and TFLopt. For the PCopt
functional, the parameters to be optimized were those of the
$\theta_{PC}(z^{PC})$ interpolation function.  For CRopt and TFLopt,
the parameters to be optimized were the coefficients of the second-order
gradient expansion, yielding
\begin{equation}
z^{CRopt} = a s^2 + b q - F_t^{W}
\end{equation}
and
\begin{equation}
F_t^{TFLopt} = \mathrm{max}\left( 1 + a s^2 + b q, F_t^{vW} \right)
\label{TFLopt}
\end{equation}
respectively. 

In addition, a new interpolation function
\begin{equation}
\theta_{\textsc{tanh}}(z) = \left\{ \mathrm{Tanh}(\left|z \right|^{-8})[1 - H(z)] + H(z) \right\}^{1/8}
\end{equation}  
\\
was used to define the new functional TANH as
\begin{equation}
F_t^{TANH} = 1 + F_t^{W} +  z^{CRopt} \theta_{\textsc{tanh}}\left( z^{CRopt} \right) \;. 
\end{equation}
The motivation is that this form imitates closely the TFLopt 
functional without having the
discontinuous derivative introduced by the $max$ function
in Eq.\ (\ref{TFLopt}). 

\begin{table}
	\caption{Error indicator $\Delta_{\alpha}+\Delta_{\alpha}^{near}$ values for the reoptimized mGGA\{a,b\} kinetic energy density functional approximations. \label{t:Dalpha2}}
	\begin{tabular}{l l r l r l r }
		       & \phantom{a} & \multicolumn{1}{c}{$a$}  & \phantom{a} & \multicolumn{1}{c}{$b$} &  \phantom{a} & \multicolumn{1}{c}{$\Delta_{\alpha}+\Delta_{\alpha}^{near}$}  \\\toprule
		PC     & \phantom{a} &   0.538900               & \phantom{a} & 3.000000                &  \phantom{a} &    0.712057                            \\
		PCopt  & \phantom{a} &   1.784720               & \phantom{a} & 0.258304                &  \phantom{a} &    0.649567                            \\
		CRloc  & \phantom{a} &  -0.275000               & \phantom{a} & 2.895000                &  \phantom{a} &    0.631376                            \\
		TFLreg & \phantom{a} &   0.000000               & \phantom{a} & 2.222222                &  \phantom{a} &    0.398936                            \\
		CRopt  & \phantom{a} &  -0.295491               & \phantom{a} & 2.615740                &  \phantom{a} &    0.383805                            \\
		TANH   & \phantom{a} &  -0.216872               & \phantom{a} & 2.528000                &  \phantom{a} &    0.365022                            \\
		TFLopt & \phantom{a} &  -0.203519               & \phantom{a} & 2.513880                &  \phantom{a} &    0.361805                            \\\toprule

	\end{tabular}
\end{table}

Parameter optimization used  \textit{Mathematica's} \texttt{NMinimize} built-in procedure \cite{mathematica}
with the Nelder-Mead method \cite{NelderMead1965}.

Table \ref{t:Dalpha2} shows $\Delta_{\alpha}+\Delta_{\alpha}^{near}$
for each of the newly reparametrized functionals as well as the
originals.  The PC, PCopt, and CRloc functionals clearly are the worst
performers.  However, even though PCopt has the second-worst error
indicator, it is not ignorable.  The reason is that it retains almost
all the asymptotic behavior of the original PC functional, and, as a
consequence, of $\alpha^{approx}$ as well. The only limiting behavior
modified by the reparametrization is that $F_t$ approaches $ 0.906485
+ F_t^{\textsc{w}} $ instead of $ 1.0 + F_t^{\textsc{w}} $ when $| q |
\rightarrow \infty$ (this is a direct consequence of $a$ being larger
than $1$).  Moreover, when $\Delta_{\alpha}$ and
$\Delta_{\alpha}^{near}$ are minimized independently, the optimized
parameters for PCopt are almost identical for each case, suggesting
balanced performance among core, valence and, tail density
regions. This is not true for any of the other KEDFs.

At this point, it is important to reiterate that good performance for the 
chosen error indicators may not translate into correspondingly
good performance in  the total non-interacting KE.
The converse also is true. Bad performers with respect to those error 
indicators may have beneficial error cancellation when
integrated, thereby yielding  a very good total non-interacting KE estimate.
Post-scf noninteracting kinetic energies listed in Table \ref{t:ts} illustrate
the point. 
  
\begin{table*}
	\caption{Post-scf noninteracting kinetic energy from various kinetic energy density functional approximations,\label{t:ts}}
	\begin{tabular}{l . . . . .}
		                & \multicolumn{1}{c}{He} & \multicolumn{1}{c}{Ne} & \multicolumn{1}{c}{Ar} & \multicolumn{1}{c}{Kr} & \multicolumn{1}{c}{Xe} \\\toprule
		PC              & 2.99305 & 129.3158 & 530.6552 & 2761.1804 & 7249.7497 \\
		PCopt           & 2.99491 & 123.5084 & 494.3828 & 2538.1368 & 6625.5351 \\
		CRloc           & 3.03627 & 126.7640 & 511.8635 & 2659.1147 & 6988.2001 \\
		CRopt           & 3.02671 & 125.9625 & 508.7109 & 2645.2305 & 6955.5975 \\
		TANH            & 2.90845 & 123.3855 & 503.7825 & 2633.6734 & 6938.1354 \\
		TFLreg          & 3.00568 & 128.6246 & 524.2290 & 2724.8688 & 7155.7830 \\		
		TFLopt          & 2.88039 & 123.1157 & 503.5572 & 2634.4597 & 6941.5470 \\\hline
		KS              & 2.86168 & 128.5471 & 526.8175 & 2752.0549 & 7232.1390 \\\toprule
	\end{tabular}
\end{table*}
Also it is interesting to note that all KEDFs listed in Table \ref{t:ts}
enforce the Weisz\"acker bound, however, $F_t^{\textsc{w}}$ is not exact
at the nuclei of elements with occupied $p$-orbitals \cite{PRB.91.035126}. 
Excluding PC and PCopt, which behave (approximately) as $1+F_t^{\textsc{w}}$
at the nuclei, it is possible that the bad performance in the total non-interacting KE
is due to the missing $p$-shell contribution, which can amount to 
12\% of the KE in the $Z \rightarrow \infty$ limit.

\section{MVS exchange functional}

We consider deorbitalizations of specific mGGAs, beginning with the  
metaGGA ``Made Very Simple'' (MVS) exchange functional \cite{MVS2015}.

The exchange energy can be written as
\begin{equation}
E_{\mathrm{x}}[n] = \int n \, \varepsilon_{\mathrm{x}}^{\mathrm{unif}} \, F_{\mathrm{x}}(s,\alpha) \; d\mathbf{r}
\end{equation}
where, $\varepsilon_{\mathrm{x}}^{\mathrm{unif}} := -(3/4)(3 n/\pi)^{1/3}$
is the uniform-electron gas exchange energy per particle and 
$F_{\mathrm{x}}(s,z,\alpha,\dots)$ is known as the enhancement factor.
The MVS exchange enhancement factor, $F_{\mathrm{x}}^{MVS}(s,\alpha)$, 
dis-entangles the $\alpha[n]$ and $s$ dependencies and respects
several constraints including the second-order gradient expansion for the
slowly varying density with coefficient $\mu = 10/81$ and the asymptotic expansion of
the exchange energy of neutral atoms. Explicitly, 
\begin{equation}
F_{\mathrm{x}}^{MVS}(s,\alpha) = \frac{1 + 0.174\,f_{\mathrm{x}}(\alpha) }{(1 + 0.0233\,s^4)^{1/8}}
\end{equation}
where the function $f_{\mathrm{x}}(\alpha)$ is given by
\begin{equation}
f_{\mathrm{x}}(\alpha) = \frac{1-\alpha}{\left[(1 - 1.6665 \, \alpha^2)^2 + 0.7438 \, \alpha^4\right]^{1/4}}
\end{equation}
The MVS mGGA exchange is paired to the modified PBE GGA correlation used
for the revTPSS functional \cite{revTPSS}.

Deorbitalized versions of the MVS mGGA exchange functional \cite{MVS2015} 
were implemented in deMon2k\cite{deMon2kv4} and in NWChem\cite{NWChem}. 
The results quoted here are from NWChem; deMon2k results are consistent.
  The NWChem calculations used  the
def2-TZVPP basis set \cite{def2TZVPP} and extra-fine grid settings. 
Details of the testing which led to use of that grid are in the section
on the SCAN mGGA below.   
Heats of formation were computed according to the established 
procedure from Curtiss {\it et al.} \cite{Gn_2}
for the 223 molecules
of the G3X/99 test set \cite{G3X}. The T96-R set \cite{T96RandT82F_1,T96RandT82F_2} 
 was used to obtain the 
optimized bond lengths statistics, and the T82-F set \cite{T96RandT82F_1,T96RandT82F_2}
for the harmonic vibrational frequencies. 

Table \ref{t:MVSdeorb} presents the results for the two variants of
each of the three deorbitalization candidates presented above.  
The PCopt and CRopt results are a striking example of an unexpected finding.
The PCopt error results are reasonably 
close to the original MVS values, while the CRopt error magnitudes
are substantially superior.
  
The difference illustrates two 
quite distinct deorbitalization objectives regarding a given mGGA 
XC functional.  
One is {\it faithful} deorbitalization,
the other is {\it best-performance} deorbitalization.  The {\it faithful} 
deorbitalization objective is to produce test-set results that are as
nearly indistinguishable as possible from those of the original mGGA.
The {\it best performance} objective is the 
deorbitalization that reduces the error magnitudes (on the original test
sets) as much as feasible below the original mGGA results. Additional 
stipulations are that the number of fitting parameters in the
deorbitalization should not be greatly in excess of the number in the
original mGGA and that the fitting not be to the test sets themselves.  

Perhaps it is no surprise that the most {\it faithful} MVS
deorbitalization follows from the PCopt KEDF, since that KEDF does not
alter the gradient expansion of $\alpha[n]$ for slowly varying
densities.  That expansion was explicitly taken into account in the
development of MVS. The very bad performance of TFLreg and TFLopt may
be the consequence of several factors. First, imposition of the
von Weizs\"acker bound through the {\it max} function can introduce
discontinuities in the potential, which in turn can lead to badly
behaved densities.  Second, TFLreg and TFLopt describe $\alpha$ very
well inside the core region of the first eighteen atoms but have the
largest deviations, among the KEDFs tested, for the valence region of
C.

As already noted, the {\it best-performance} deorbitalization was
unanticipated.  The finding regarding MVS is unequivocal however.  On
these test sets at least, MVS is improved by conversion to a
Laplacian-level functional, mGGA-L, rather than the conventional mGGA
form in which it was developed. On the test sets considered, MVS-L
delivers performance quite similar to the highly-sophisticated SCAN
mGGA functional.
 
\begin{table*}
	\caption{Performance of the deorbitalized versions of the MVS exchange-correlation functional. Heat of formation errors in kcal/mol, bond length errors in {\AA}ngstrom, frequency errors in cm$^{-1}$. }\label{t:MVSdeorb}
	\begin{tabular}{l l c  . . . . . . . }
		                                    &     & \phantom{cc} &  \multicolumn{1}{c}{PC} & \multicolumn{1}{c}{PCopt} & \multicolumn{1}{c}{TFLreg} & \multicolumn{1}{c}{TFLopt} & \multicolumn{1}{c}{CRloc} & \multicolumn{1}{c}{CRopt} & \multicolumn{1}{c}{MVS}  \\\toprule
		\multirow{2}{*}{Heats of formation} & ME  & \phantom{cc} & 24.00                   & -15.37                    & 18.27                      &  19.18                     & 2.71                      & 2.89                      & -17.33   \\
		                                    & MAE & \phantom{cc} & 25.53                   &  15.94                    & 19.09                      &  19.91                     & 7.53                      & 6.20                      &  18.34   \\\hline
		\multirow{2}{*}{Bonds}              & ME  & \phantom{cc} & 0.0069                  &  -0.0025                  & 0.0092                     &  0.0072                    & 0.0025                    & 0.0049                    & -0.0016  \\
		                                    & MAE & \phantom{cc} & 0.0137                  &   0.0127                  & 0.0139                     &  0.0139                    & 0.0121                    & 0.0130                    &  0.0139  \\\hline
		\multirow{2}{*}{Frequencies}        & ME  & \phantom{cc} &  2.9                    &  39.3                     &  9.6                       &  20.0                      & 25.3                      & 28.7                      &  46.2    \\
		                                    & MAE & \phantom{cc} & 29.4                    &  46.0                     & 34.7                       &  37.7                      & 37.0                      & 42.6                      &  52.0    \\\toprule
	\end{tabular}
\end{table*}

\section{TPSS exchange-correlation functional}

Of the three mGGAs considered for deorbitalization, 
the TPSS exchange-correlation functional \cite{TPSSa} is the most challenging
 case.
It has orbital dependence in both exchange and correlation terms. Moreover,
the TPSS exchange enhancement factor, $F_{\mathrm{x}}^{TPSS}(s,z,\alpha)$,
has the additional complication of being dependent upon 
 two orbital-dependent dimensionless ratios, $z[n]$ and $\alpha[n]$.
Such complication does not occur in either MVS (compare above) or SCAN 
(compare below) enhancement factors. 
The TPSS exchange enhancement factor is given by
\begin{equation}
F_{\mathrm{x}}^{TPSS}(s,z,\alpha) = 1 + \kappa -\frac{\kappa^2}{\kappa + x(s,z,\alpha)}
\end{equation}
where
\begin{eqnarray}
x &=&  \left\lbrace \left[ \mu_{\textsc{ge}} + c\frac{z^2}{\left( 1+z^2 \right)^2 } \right] s^2 + \frac{146}{2025}\tilde{q}_b^2  \right. \nonumber \\
  &-& \left. \frac{73}{405}\tilde{q}_b \sqrt{ \frac{1}{2} \left( \frac{3}{5}z \right)^2 + \frac{1}{2}s^4} + \frac{\mu_{\textsc{ge}}^2}{\kappa}s^4 \right. \nonumber \\
  &+& \left. 2\sqrt{e}\mu_{\textsc{ge}} \left( \frac{3}{5}z\right)^2 + e\mu_{\textsc{pbe}}s^6 \right\rbrace \Bigg/ \left( 1+\sqrt{e} s^2 \right)^2
\end{eqnarray}
and
\begin{equation}
\tilde{q}_b = \frac{\tfrac{9}{20}(\alpha - 1)}{\left[ 1+b \alpha (\alpha - 1)\right]^{1/2}} + \frac{2}{3}s^2
\end{equation}
The constants $\kappa = 0.804$, $\mu_{\textsc{ge}} = 10/81$,
$\mu_{\textsc{pbe}} = 0.21951$, $b=0.40$, $c=1.59096$ and $e=1.537$,
were fixed by imposition of several conditions.

TPSS correlation has slightly simpler
orbital-dependence in that it depends only upon the 
dimensionless ratio, $z$.  It can be written as
\begin{equation}
E_c[n] = \int n \, \epsilon_c^{\textsc{tpss}} \; d\mathbf{r}
\end{equation}
where
\begin{equation}
\epsilon_c^{\textsc{tpss}} = \epsilon_c^{rev\textsc{pkzb}} \left[ 1 + 2.8 z^3 \epsilon_c^{rev\textsc{pkzb}} \right]
\end{equation}
The revised PKZB correlation energy $\epsilon_c^{rev\textsc{pkzb}}$ is an mGGA itself
which is given by
\begin{eqnarray}
\epsilon_c^{rev\textsc{pkzb}} &=& \epsilon_c^{\textsc{pbe}}\left[ 1 + C(\zeta,\xi) z^2\right] \nonumber \\
                     &-& \left[ 1 + C(\zeta,\xi)\right] z^2 \sum_\sigma \frac{n_\sigma}{n} \tilde{\epsilon}_c^{\sigma}
\end{eqnarray}
with
\begin{equation}
C(\zeta,\xi) = \frac{0.53 + 0.87\zeta^2 + 0.50\zeta^4 + 2.26\zeta^6}{\left\lbrace 1 + \xi^2 \left[ \left( 1 + \zeta\right)^{-4/3} + \left( 1 - \zeta \right)^{-4/3} \right]/2  \right\rbrace^4} \; ,
\end{equation}
\be
\zeta = \frac{n_{\uparrow} - n_{_\downarrow}}{n} \;\; , \;\; \xi = \frac{\grad \zeta}{2 (3\pi^2 n)^{1/3}}
\ee
and
\begin{equation}
\tilde{\epsilon}_c^{\sigma} = \mathrm{max} \left[ \epsilon_c^{\textsc{pbe}}(n_{\uparrow},n_{\downarrow},\grad n_{\uparrow},\grad n_{\downarrow}) , \epsilon_c^{\textsc{pbe}}(n_\sigma,0,\grad n_\sigma,0) \right]
\end{equation}
Here $\epsilon_c^{\textsc{pbe}}$ is the PBE (GGA) correlation energy per particle
\cite{PBE}.

Table \ref{t:TPSSdeorb} presents the results for the deorbitalized
TPSS variants.  It did not prove possible to achieve a {\it
  best-performance} case with any of the KEDF candidates we
examined.  The most nearly {\it faithful} case was obtained through the TFLreg
KEDF. That KEDF provides the most balanced behavior for describing both $z[n]$
and $\alpha[n]$. Even that deorbitalization is not entirely 
successful, in that all of its MAE values are worse than those from 
the original TPSS.  Nevertheless TPSS-L (with TFLreg) may be good enough
to be useful in the simulation context.

\begin{table*}
	\caption{As in Table \ref{t:MVSdeorb} for the deorbitalized versions of the TPSS exchange-correlation functional.}\label{t:TPSSdeorb}
	\begin{tabular}{l l c  . . . . . . . }
		                                    &     & \phantom{cc} &  \multicolumn{1}{c}{PC} & \multicolumn{1}{c}{PCopt} & \multicolumn{1}{c}{TFLreg} & \multicolumn{1}{c}{TFLopt} & \multicolumn{1}{c}{CRloc} & \multicolumn{1}{c}{CRopt} & \multicolumn{1}{c}{TPSS}  \\\toprule
		\multirow{2}{*}{Heats of formation} & ME  & \phantom{cc} & 10.13                   &  13.51                    &  1.14                      &   7.45                     & 12.65                     & 9.15                      & -4.37   \\
		                                    & MAE & \phantom{cc} & 12.29                   &  15.61                    &  6.24                      &  10.86                     & 15.05                     & 11.89                     &  5.28   \\\hline
		\multirow{2}{*}{Bonds}              & ME  & \phantom{cc} & 0.0212                  &  0.0215                   &  0.0209                    &  0.0214                    &  0.0224                   & 0.0217                    &  0.0134 \\
		                                    & MAE & \phantom{cc} & 0.0215                  &  0.0217                   &  0.0209                    &  0.0217                    & 0.0224                    & 0.0218                    &  0.0156 \\\hline
		\multirow{2}{*}{Frequencies}        & ME  & \phantom{cc} & -42.7                   &  -44.5                    &  -42.5                     &  -44.2                     & -46.5                     & -45.2                     &  -18.3  \\
		                                    & MAE & \phantom{cc} & 49.6                    &  50.2                     &  47.5                      &  49.5                      & 51.4                      & 50.0                      &  31.2   \\\toprule
	\end{tabular}
\end{table*}

\section{SCAN exchange-correlation functional}

The strongly constrained and appropriately normed (SCAN)
exchange-correlation functional is said to provide the best 
overall performance
among all non-empirical mGGAs developed so far. Its orbital-dependence
comes from $\alpha[n]$ alone for both exchange and correlation, making it 
a good deorbitalization candidate.

The SCAN exchange enhancement factor is given by
\begin{eqnarray}
F_\mathrm{x}^{\textsc{scan}}(s,\alpha) &=& \left\lbrace h_x^1(s,\alpha) \nonumber \right. \\
                                       &+& \left. f_x(\alpha) \left[ 1.174 - h_x^1(s,\alpha) \right] \right\rbrace g_x(s)
\end{eqnarray}
with
\begin{equation}
g_x(s) = 1 - e^{-a_1/\sqrt{s}}
\end{equation}
and
\begin{equation}
f_x(\alpha) = e^{-c_{1x}\alpha/(1-\alpha)}\theta(1-\alpha) - d_x e^{c_{2x}/(1-\alpha)}\theta(\alpha-1)
\end{equation}
The remaining function $h_x^1(s,\alpha)$ is an approximate resummation
of the fourth-order gradient expansion for exchange:
\begin{equation}
h_x^1(s,\alpha) = 1 + \frac{k_1 x}{k_1 + x}
\end{equation}
where
\begin{eqnarray}
x &=& \mu_{\textsc{ge}} s^2 \left[ 1 + \frac{b_4 s^2}{\mu_{\textsc{ge}}}e^{-b_4 s^2/\mu_{\textsc{ge}}} \right] \nonumber \\  
  &+& \left[ b_1 s^2 + b_2 (1-\alpha) e^{-b_3(1-\alpha)^2}\right]^2
\end{eqnarray}

The constants $a_1 = 4.9479$, $\mu_{\textsc{ge}} = 10/81$, $b_2 = \sqrt{5913/405000}$, $b_1 = (511/13500)/(2b_1)$,
$b_3 = 0.5$, $b_4 = \mu_{\textsc{ge}}^2/k_1 - 1606/18225 - b_1^2$, $c_{1x} = 0.667$, $c_{2x} = 0.8$,
$d_x = 1.24$ and $k_1 = 0.065$, were determined by 
imposition of known constraints or norms.

The correlation part of SCAN depends on $\alpha[n]$
only through
\begin{equation}
f_c(\alpha) = e^{-c_{1c}\alpha/(1-\alpha)}\theta(1-\alpha) - d_c e^{c_{2c}/(1-\alpha)}\theta(\alpha-1)
\end{equation}
where $c_{1c}=0.64$, $d_c=0.7$ and $c_{2c}=1.5$. The interpolation function, $f_c$,
interpolates between two revised PBE correlation energies per particle, $\epsilon_c^0$ and $\epsilon_c^1$, 
valid for $\alpha=0$ and $\alpha = 1$, respectively:
\be
E_c^{SCAN} = \int \; d\mathbf{r} \;n \Big[ \epsilon_c^1 + f_c(\alpha) \left( \epsilon_c^0 - \epsilon_c^1 \right) \Big]
\ee

Results from the deorbitalization of SCAN are shown in Table
\ref{t:SCANdeorb}. One sees that the {\it faithful} case is
essentially achieved by the PCopt functional, with very similar
performance to original SCAN for thermochemistry and bond lengths and
slightly better performance for harmonic vibrational frequencies than
the orbital-dependent SCAN.  No example of a {\it best performance}
deorbitalization occurred within the range of candidate KEDFs.  The
{\it faithful} deorbitalization is remarkable nevertheless: SCAN-L
with PCopt is just about as good as SCAN on these standard test sets.

A subtlety is involved in the computational implementations of SCAN
that we have studied. The exchange contribution is computed using the
spin-scaling relation, $E_x[n_{\uparrow},n_{\downarrow}] = \tfrac{1}{2}\left(E_x[2n_{\uparrow}]+E_x[2n_{\downarrow}]\right)$,
which means that
\be
\alpha_\sigma = \frac{2 t_s^\sigma - t_W[2n_\sigma]}{t_{TF}[2n_\sigma]}
\ee
However, $\alpha$ in the correlation contribution is computed as (see
Supplemental Material of Ref. \onlinecite{SCAN2015})
\begin{eqnarray}
\alpha &=& \frac{t_s - t_W[n]}{\tfrac{1}{2}\left( (1+\zeta)^{5/3} + (1-\zeta)^{5/3}\right)t_{TF}[n]} \nonumber\\
       &=& \frac{\tfrac{1}{2}\sum_{\sigma}2t_s^{\sigma} - t_W[n]}{\tfrac{1}{2}\sum_{\sigma} t_{TF}[2n_\sigma]}
\end{eqnarray}
Note that in the latter case, $\alpha$ uses the
spin-scaling relation except for the Weisz\"acker piece (this mismatch
was confirmed by reproducing Table SIV from the Supplemental Material of
Ref. \onlinecite{SCAN2015}). As a consequence, a deorbitalization
with higher fidelity might be obtained if the KEDF were to be 
optimized separately for exchange and correlation.

\begin{table*}
	\caption{As in Table \ref{t:MVSdeorb} for the deorbitalized versions of the SCAN exchange-correlation functional.}\label{t:SCANdeorb}
	\begin{tabular}{l l c  . . . . . . . }
		                                    &     & \phantom{cc} &  \multicolumn{1}{c}{PC} & \multicolumn{1}{c}{PCopt} & \multicolumn{1}{c}{TFLreg} & \multicolumn{1}{c}{TFLopt} & \multicolumn{1}{c}{CRloc} & \multicolumn{1}{c}{CRopt} & \multicolumn{1}{c}{SCAN}  \\\toprule
		\multirow{2}{*}{Heats of formation} & ME  & \phantom{cc} &  17.57                  &  2.11                     &  56.91                     &  63.60                     &  14.82                    & 7.75                      & -3.62   \\
		                                    & MAE & \phantom{cc} &  19.99                  &  5.67                     &  57.23                     &  63.97                     &  16.70                    & 14.81                     &  5.12   \\\hline
		\multirow{2}{*}{Bonds}              & ME  & \phantom{cc} &  0.0158                 &  0.0073                   &  0.0198                    &  0.0190                    &  0.0140                   & 0.0168                    &  0.0035 \\
		                                    & MAE & \phantom{cc} &  0.0189                 &  0.0105                   &  0.0220                    &  0.0221                    &  0.0155                   & 0.0181                    &  0.0089 \\\hline
		\multirow{2}{*}{Frequencies}        & ME  & \phantom{cc} &  -38.5                  & -11.7                     &  -47.8                     &  -43.2                     &  -27.7                    & -29.8                     &  15.3    \\
		                                    & MAE & \phantom{cc} &  49.3                   &  28.7                     &   54.5                     &   51.2                     &  38.5                     &  38.6                     &  31.9    \\\toprule
	\end{tabular}
\end{table*}

Exploration of SCAN-L (PCopt) illustrates issues of behaviors 
with respect to changes in numerical integration grid or basis set
size which we encountered repeatedly.    
Tables \ref{t:grid} and \ref{t:basis} compare the results from original
SCAN and the SCAN-L (PCopt) functional for grid and basis set changes, 
respectively. 

The Table \ref{t:grid} results were obtained with the def2-TZVPP basis 
set for three levels of grid quality  
pre-defined  in NWChem. Use of the default grid quality, {\it medium}, led to
self-consistent field (scf) convergence issues that rendered the {\it medium}
option clearly inappropriate for both original SCAN and SCAN-L. The
{\it fine} quality grid produces well-converged results for all test sets 
in the case of original SCAN.  It also does well for  
the thermochemistry and bond length sets with SCAN-L, but
does not work for the vibrational frequencies set.  The same behavior 
was observed for the deorbitalized versions of MVS and TPSS: the {\it xfine}
grid quality was needed in order to ensure converged vibrational 
frequencies with the Laplacian-dependent functionals.

\begin{table*}
	\caption{Grid size sensitivity for SCAN and SCAN-L (PCopt). \label{t:grid}}
	\begin{tabular}{l l . . . l . . .}
		                                    &     & \multicolumn{3}{c}{SCAN}                                                          & \phantom{c} & \multicolumn{3}{c}{SCAN-L}                                                    \\
		                                    &     & \multicolumn{1}{c}{medium} & \multicolumn{1}{c}{fine} & \multicolumn{1}{c}{xfine} & \phantom{c} & \multicolumn{1}{c}{medium} & \multicolumn{1}{c}{fine} & \multicolumn{1}{c}{xfine} \\\toprule
		\multirow{2}{*}{Heats of formation} & ME  & -4.04                      & -3.61                    & -3.62                     &             & 1.45                       &  2.14                    & 2.11                      \\ 
		                                    & MAE & 5.76                       & 5.13                     &  5.12                     &             & 5.90                       &  5.71                    & 5.67                      \\\hline
		\multirow{2}{*}{Bonds}              & ME  & 0.0035                     & 0.0036                   &  0.0035                   &             & 0.0074                     &  0.0072                  & 0.0073                    \\
		                                    & MAE & 0.0089                     & 0.0090                   &  0.0089                   &             & 0.0108                     &  0.0106                  & 0.0105                    \\\hline
		\multirow{2}{*}{Frequencies}        & ME  & 25.1                       & 15.2                     &  15.3                     &             & -9.3                       &  -3.0                    & -11.7                     \\
		                                    & MAE & 42.9                       & 33.5                     &  31.9                     &             & 41.3                       &  33.4                    & 28.7                      \\\toprule
	\end{tabular}
\end{table*}

\begin{table*}
	\caption{Basis set sensitivity for SCAN and SCAN-L (PCopt). Double (``svp''), triple (``tzvpp''), quadruple (``qzvpp'') and complete basis set 
extrapolation (``cbs'') error values. \label{t:basis}}
	\begin{tabular}{l l . . . . l . . . .}
		                                    &     & \multicolumn{4}{c}{SCAN}                                                                                       & \phantom{c} & \multicolumn{4}{c}{SCAN-L}                                                    \\
		                                    &     & \multicolumn{1}{c}{svp}    & \multicolumn{1}{c}{tzvpp} & \multicolumn{1}{c}{qzvpp} & \multicolumn{1}{c}{cbs}   & \phantom{c} & \multicolumn{1}{c}{svp} & \multicolumn{1}{c}{tzvpp} & \multicolumn{1}{c}{qzvpp} & \multicolumn{1}{c}{cbs}   \\\toprule
		\multirow{2}{*}{Heats of formation} & ME  &  -7.06                     &  -3.62                    &  -4.09                    & -4.15                     &             & -6.44                   & 2.11                      & 2.63                      & 2.71                      \\ 
		                                    & MAE &  11.48                     &   5.12                    &   5.32                    &  5.35                     &             & 10.67                   & 5.67                      & 5.67                      & 5.68                      \\\hline
		\multirow{2}{*}{Bonds}              & ME  &  0.0129                    &   0.0035                  &   0.0018                  &                           &             & 0.0150                  & 0.0073                    & 0.0057                    &                           \\
		                                    & MAE &  0.0189                    &   0.0089                  &   0.0081                  &                           &             & 0.0193                  & 0.0105                    & 0.0095                    &                           \\\hline
		\multirow{2}{*}{Frequencies}        & ME  &  21.9                      &   15.3                    &   16.5                    &                           &             & 00.0                    & -11.7                     & -9.6                      &                           \\
		                                    & MAE &  41.7                      &   31.9                    &   32.3                    &                           &             & 34.0                    & 28.7                      & 30.8                      &                           \\\toprule
	\end{tabular}                                                                                                                                               
\end{table*}

On the other hand, SCAN and SCAN-L both have the same behavior with 
respect to increasing the basis set cardinality.  They yield  
oscillatory behavior in the heats of formation and vibrational 
frequency errors along with a steady 
shortening of bond length errors when changing from double- to triple- to
quadruple-zeta basis sets. Also shown in Table \ref{t:basis} are
extrapolations to the complete basis-set (CBS) limit obtained with
a linear extrapolation of the scf energy \cite{PeterssonCBS,NeeseCBS}. Those
extrapolations indicate that the def2-TZVPP results are about 0.5 kcal/mol
from the basis set limit for both original SCAN and SCAN-L.

\section{Conclusions}

We have shown that it is possible to reduce the complexity yet retain
the quality of some mGGAs 
by deorbitalization.  To remove the orbital-dependence,
a kinetic energy density functional which reproduces as closely as
possible the relevant orbital-dependent dimensionless variable must be
selected. We recommend the PC, CRloc, and TFLreg KEDFs, along with 
their respective reparametrizations, as the most-likely-to-succeed ones.

We emphasize that none of the parametrization is to experimental
data or to data from outside the domain of DFT.  Rather it is
parametrization of approximate KEDFs to deliver (post-scf) 
values of the non-interacting KE of the first 18 
atoms as close possible to the KE values obtained from the
(post-scf) orbitals on those atoms. 

We have delineated the difference between {\it faithful} and {\it best 
performing} deorbitalizations.  In the {\it faithful} case, the
result is a Laplacian-level functional that recovers essentially the
same properties for finite systems as the original. Because the
Laplacian-level functional obviates the use of generalized-KS
equations, in principle it is computationally more efficient 
than calculation with 
the original mGGA. Moreover, the deorbitalized functional also gives an
approximate rendition of the local potential that the OEP would give 
for the original
functional.  We have not explored exploitation of this last fact.

{\it Best performance} deorbitalization was unexpected.  
When it can be achieved, the resulting Laplacian-level functional 
actually combines better numerical results with a
less complicated, less computationally demanding functional than the original 
mGGA.  This outcome is an illustration that  
the putative relationship between the rung of the Perdew-Schmidt
Jacob's ladder of functional complexity and performance is not
as direct and unambiguous as seems to be widely believed.  
The computational complexity advantage gained by {\it best performance} 
deorbitalization is the same as for the {\it faithful} case.

Functionals with orbital-dependences from two dimensionless ratios,
such as TPSS, pose serious challenges to the deorbitalization
scheme presented here.  Because of that TPSS complexity, other deorbitalization
approaches might yield better results than those obtained here. For
example, it would be possible to reparametrize the KEDFs to minimize
an error indicator based on $z[n]$ or reparametrize using error
measures based on both $\alpha[n]$ and $z[n]$ concurrently.  Different
deorbitalization KEDFs might be used for exchange and 
correlation. These routes were not explored since one of the
objectives of this initial study was to keep the deorbitalization
strategy as simple as possible.

On the other hand, MVS and SCAN, which depend only on $\alpha[n]$, 
are demonstrably very good candidates for deorbitalization. In particular, 
SCAN-L, the PCopt deorbitalized SCAN, seems to be the most
accurate XC functional presently available for use in either
true KS (not generalized KS) calculations or in orbital-free DFT 
calculations.   

Although the proposed deorbitalized functionals are
Laplacian-dependent, their computational stability is comparable to
that of standard mGGA functionals. Roughly the same number of
self-consistent field cycles, as well as geometry optimization steps,
were needed to converge the standard and de-deorbitalized versions of
three functionals tested.  The one exception to this general
comparability is the requirement of extra-fine grids. 

The main drawback of the Laplacian-dependent functionals for
molecular calculations as we have done them so far comes 
from computing the Laplacian of the density, as well
as the matrix elements from the exchange-correlation potential.  
These cause a noticeable performance impact. That impact arises from the
need for higher-order derivatives of the basis functions.  Our tests were 
implemented without concern for calculational efficiency, that is to
say, in the simplest way possible.  Thus we used extant coding
where possible.  In NWChem, for example, this means that we compute the 
non-redundant part of the hessian matrix for each basis function, 
instead of only the three relevant second-order derivatives. If 
deorbitalization were to be accepted as part of the computational
strategy, this impact could be ameliorated substantially 
by writing optimized code which computes only the
relevant derivatives (trace of the hessian).  

We note that corresponding performance degradation is not expected to
occur in periodic system computations because the Laplacian terms can
be computed with no significant cost in Fourier space.  We are 
refining and testing the present deorbitalization schemes on
such systems at this writing.  \\

\begin{acknowledgments}
	We acknowledge, with hearty thanks, the initial deorbitalization 
	explorations of Debajit Chakraborty.  
	This 
	work was supported  by U.S.\ National Science Foundation 
	grant DMR-1515307.
	
\end{acknowledgments}

\end{document}